\documentstyle[twoside,fleqn,espcrc2,epsf]{article}

\newcommand{\AmS}{{\protect\the\textfont2
  A\kern-.1667em\lower.5ex\hbox{M}\kern-.125emS}}

\newlength{\numlen}

\newcommand{\h}{{\hspace{0.5 cm}}}

\newcommand{\tr}{{\rm Tr\,}}
\newcommand{\nr}[1]{(\ref{#1})}
\newcommand{\fr}[2]{{\frac{#1}{#2}}}

\newcommand{\la}[1]{\label{#1}}

\newcommand{\be}{\begin{equation}}
\newcommand{\ee}{\end{equation}}
\newcommand{\ba}{\begin{eqnarray}}
\newcommand{\ea}{\end{eqnarray}}
\newcommand{\bi}{\begin{itemize}}
\newcommand{\ei}{\end{itemize}}

\def\lsi{\raise0.3ex
\hbox{$<$\kern-0.75em\raise-1.1ex\hbox{$\sim$}}}
\def\gsi{\raise0.3ex
\hbox{$>$\kern-0.75em\raise-1.1ex\hbox{$\sim$}}}

\newcommand{\gsim}{\mathop{\gsi}}

\newcommand{\half}{\mbox{${1\over2}$}}

\newcommand{\eq}{eq.\,}

\newcommand{\etal}{{\em et al}}

\newcommand{\fig}{Fig.~}

\title{% 
The Ising model universality of the electroweak theory%
\thanks{Presented by K. Rummukainen at the conference LATTICE '98,
Boulder, Colorado, July 1998.}%
}

\author{%
K. Kajantie,$\!$\address{Theory Division, CERN, CH-1211 Geneva 23, 
Switzerland}%
$^,$\address{Department of Physics, P.O.Box 9,
00014 University of Helsinki, Finland}
M. Laine,$\!^{\rm a,b}$
K. Rummukainen,$\!$\address{Nordita, Blegdamsvej 17, DK-2100 Copenhagen \O,
Denmark}
M. Shaposhnikov$^{\rm a}$ and
M. Tsypin\address{Department of Theoretical Physics,
Lebedev Physical Institute, 117924 Moscow, Russia}
\hfill\raisebox{22mm}[0mm][0mm]{\makebox[0mm][r]{\large NORDITA-98/44HE}}%
\raisebox{17mm}[0mm][0mm]{\makebox[0mm][r]{\large September 1998}}
}
\begin{document}

\begin{abstract}
Lattice simulations have shown that the first order electroweak phase
transition turns into a regular cross-over at a critical Higgs mass
$m_{H,c}$. We have developed a method which enables us to make a
detailed investigation of the critical properties of the electroweak
theory at $m_{H,c}$. We find that the transition falls into the 3d Ising
universality class. The continuum limit extrapolation of the critical
Higgs mass is $m_{H,c} = 72(2)$\,GeV, which implies that there is no
electroweak phase transition in the Standard Model.
\vspace*{-3mm}
\end{abstract}

\maketitle
\thispagestyle{empty}

%\vspace*{-50mm}

%\vspace*{45mm}

The Standard Model (SM) finite temperature phase transition has been
studied in great detail with lattice Monte Carlo simulations.  The
transition is of the first order at small Higgs masses, but it has been
found to turn into a regular cross-over when $m_H \gsim 75$\,GeV\@
\cite{isthere,karsch,gurtler,4d}.  A second order transition appears
at the endpoint of the first order transition line, and the
macroscopic behaviour of the system is determined by the {\em
universal\,} properties of the endpoint.  While the location of the
endpoint and the mass spectrum near it have been studied before, the
critical properties of the endpoint itself have not been resolved so
far.  In this talk we show that the universality class of the SM
endpoint is of the 3d
Ising type.  A full description of this work can be found
in ref.~\cite{endpoint}.

At high temperatures, the static properties of the SM and many 
of its extensions, 
can be accurately described with an effective 3d SU(2) gauge + Higgs
theory \cite{generic}:
\be
{\cal L} = 
\fr14 F^a_{ij}F^a_{ij}+|D_i\phi|^2 +
               m_3^2|\phi|^2+
               \lambda_3 |\phi|^4.
\label{action}
\ee
The theory is fixed by the dimensionful gauge coupling $g_3^2$ and by the
ratios 
\be
x=\lambda_3/g_3^2, \h y=m_3^2(\mu)/g_3^4\,,
\ee
where $m_3^2(\mu)$ is the renormalized mass parameter.
We have omitted the U(1) sector of the SM; this is justified,
since the U(1) gauge boson remains massless at any temperature and
does not affect the transition qualitatively.

%The 3D theory is latticized in a standard way; for details,
%we refer to ref.~\cite{endpoint} and references therein.

The lattice action in standard formalism is
\begin{eqnarray}
S&=& \beta_G \sum_{x;i,j} (1-\half \tr P_{ij}) \nonumber \\
 &-& \beta_H \sum_{x;i}
\half\tr\Phi^\dagger(x)U_i(x)\Phi(x+i)
\nonumber \\
 &+& \sum_x \left[
\half\tr\Phi^\dagger\Phi + \beta_R
 \bigl( \half\tr\Phi^\dagger\Phi-1 \bigr)^2 \right]
\la{latticeaction} \\
&\equiv& S_G+S_{\rm hopping}+S_{\phi^2}+S_{(\phi^2-1)^2}.
\nonumber
\end{eqnarray}
The universal behaviour does not depend on the lattice spacing,
which we keep fixed at $a \equiv 4/(g_3^2 \beta_G) = 4/(5g_3^2)$.
For the full lattice $\leftrightarrow$ continuum relations, see
\cite{endpoint} and references therein.

%%%%%%%%%%%%%%%%%%%%%%%%%%%%%%%%% FIGURE
\begin{figure}[bt]

\vspace*{-3mm}
\epsfxsize=7.7cm\epsfbox{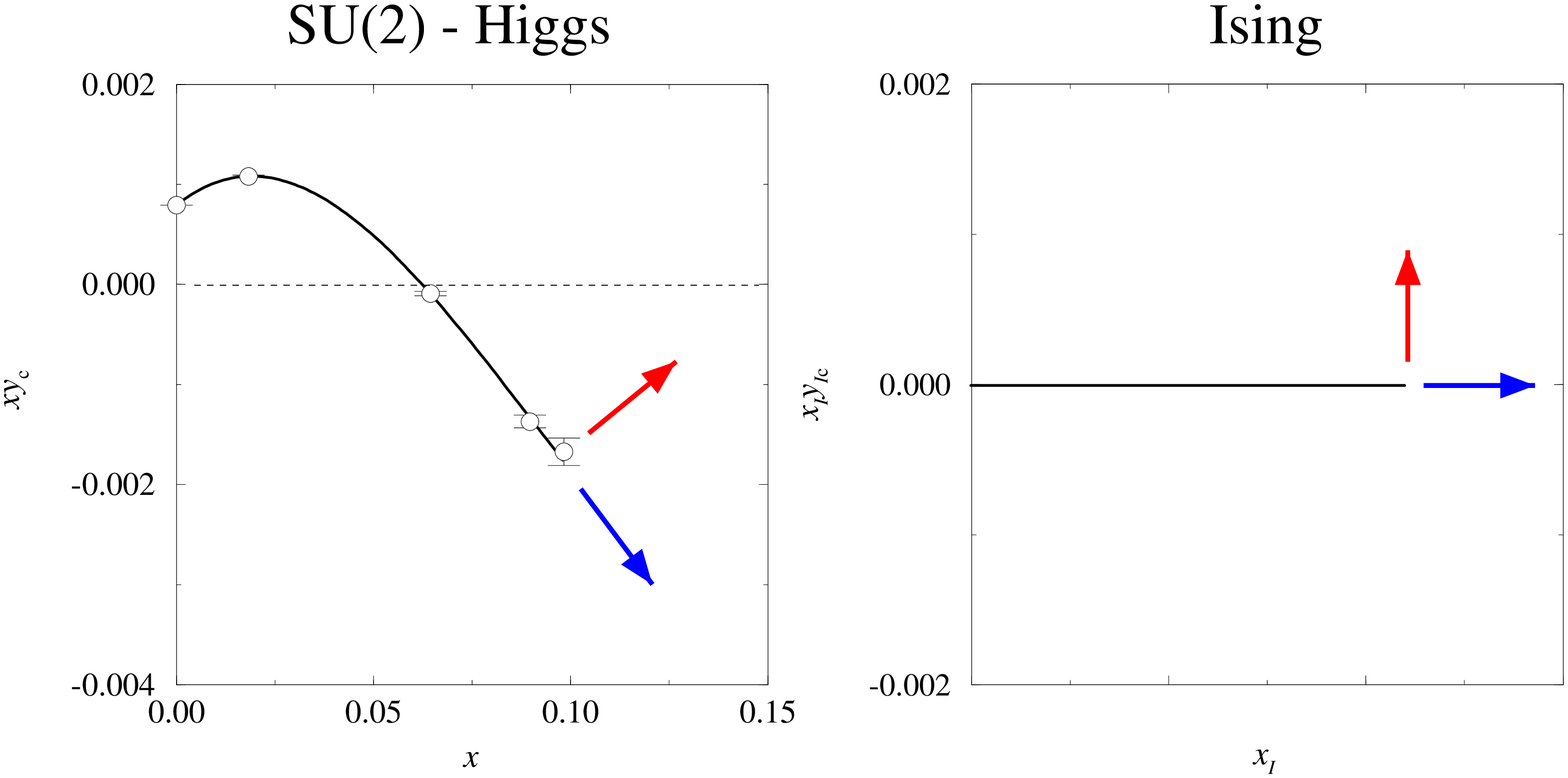}

\vspace*{-3cm}
\caption[a]{
The phase diagrams of the SU(2)+Higgs (left) and
the Ising (right) models.}
\la{fig:phasediag}
\end{figure}
%%%%%%%%%%%%%%%%%%%%%%%%%%%%%%%%%%%%

%%%%%%%%%%%%%%%%%%%%%%%%%%%%%%%%% FIGURE
\begin{figure}[t]

\vspace{2mm}
\centerline{\epsfxsize=5cm \epsffile{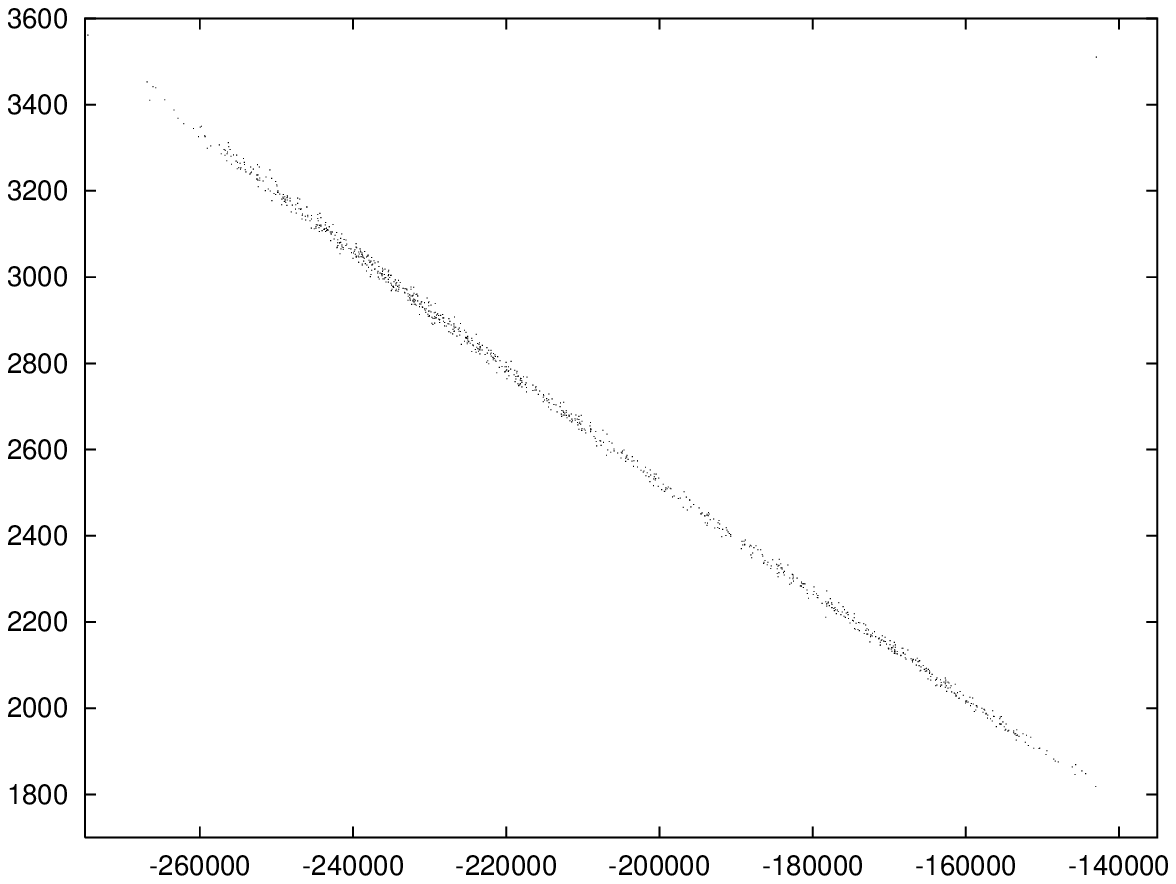}}

\centerline{\epsfxsize=5cm \epsffile{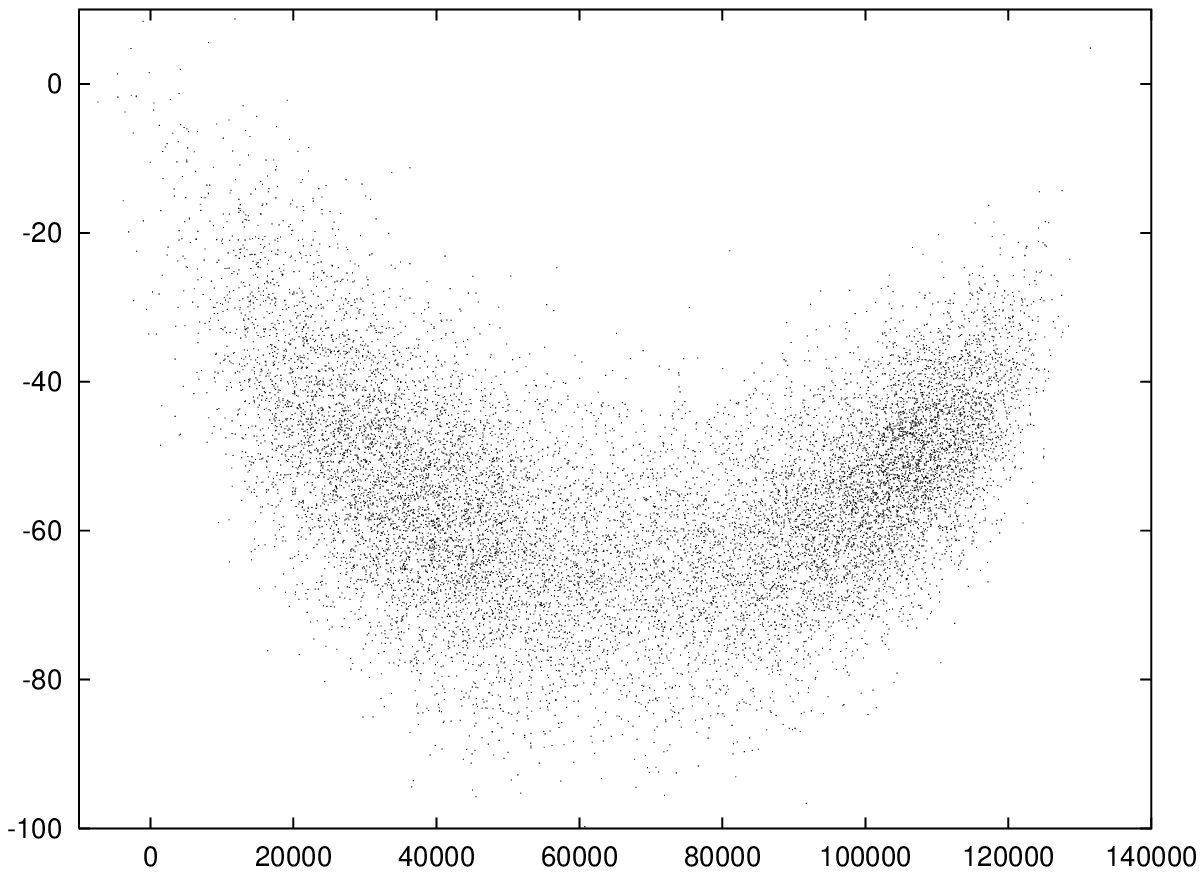}}

\centerline{\epsfxsize=5cm \epsffile{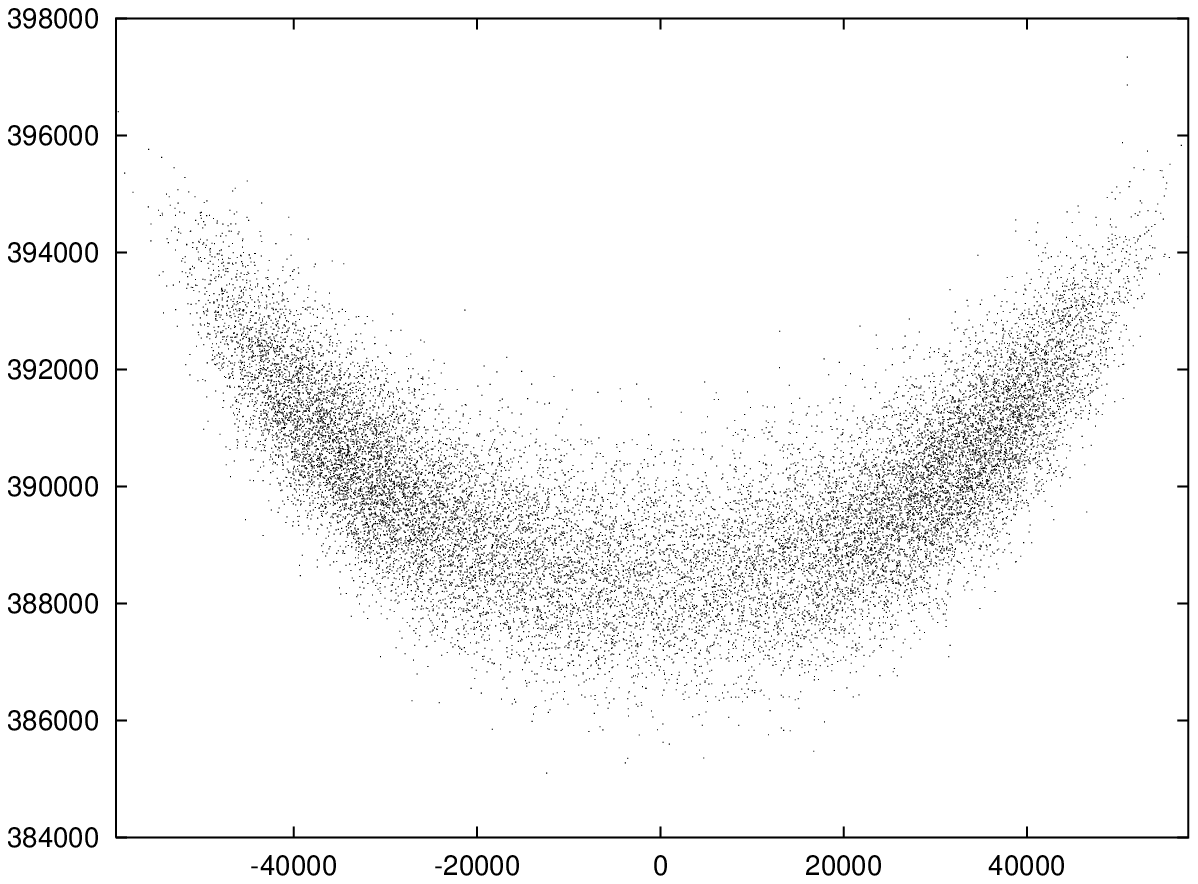}}

\vspace{-5mm}
\caption[a]{ {\em Top:} A density plot of the 3d SU(2)+Higgs model at
the critical point, shown on the $S_{(\phi^2-1)^2}$ vs.\ $S_{\rm
hopping}$ plane.  {\em Middle:} The same as above, after a shift and
a rotation.  {\em Bottom:} A density plot of the
3d Ising model on the energy vs. magnetization plane.\la{fig:density}}
\vspace{-2mm}
\end{figure}

%%%%%%%%%%%%%%%%%%%%%%%%%%%%%%%%%%%%

The phase diagram of the SU(2)+Higgs theory is shown in
\fig\ref{fig:phasediag}.  What kind of critical
behaviour can one expect?  Formally, the Higgs field
has ${\rm SU}(2)_{\rm gauge}\times {\rm SU}(2)_{\rm custodial}$
symmetry, but this remains unbroken at all temperatures.  Indeed, the
mass spectrum of the system has been investigated in detail both above
and below the critical point, and only one scalar excitation (which
couples to $\phi^\dagger\phi$) becomes light in its neighbourhood.
Thus, one would expect Ising-type universality, but also mean 
field-type or multicritical behaviour is, in principle, possible.

Adopting now Ising-model terminology, let us call the two
critical directions in \fig\ref{fig:phasediag} the magnetization
like ($M$; perpendicular to the transition line) and the
energy like ($E$; along the transition line).
Due to the lack of an exact order parameter, the mapping of the $M$-like
and $E$-like directions of the SU(2)+Higgs model
to the Ising model is non-trivial.  This is
illustrated in \fig\ref{fig:density}, where the probability
density at the critical point is plotted on the $(S_{\rm
hopping},S_{(\phi^2-1)^2})$-plane.  Only after a suitable rotation of
the axes is the striking similarity with the Ising model revealed.
This rotation closely corresponds to the rotation of the directions
shown in the phase diagrams in \fig\ref{fig:phasediag}.

However, there is no reason to restrict ourselves only to 
the two observables $S_{\rm
hopping}$ and $S_{(\phi^2-1)^2}$. Any number of operators can
contribute to the true $M$-like and $E$-like directions.  In order to
improve on the projection, it is important to consider a
large number of operators.  Our method works as follows:

\smallskip
%\noindent
(a) Locate the infinite volume critical point (for details,
see ref.~\cite{endpoint}), where all of the subsequent analysis
is performed.

%\noindent
(b) Using several volumes, measure the fluctuation matrix $M_{ij} =
\langle s_i s_j \rangle$, $ s_i \equiv S_i - \langle S_i\rangle$.
We used up to 6 operators: those in \eq\nr{latticeaction}, together
with the operators ($V = \Phi/|\Phi|$)
\[
S_R=\sum_x |\Phi|\,,\,\,\,
S_L=\sum_{x,i} \half\tr V^\dagger(x)U_i(x)V(x+i).
\]

%\noindent
(c) Calculate the eigenvalues $\lambda_\alpha$ and -vectors
$V_\alpha$ of $M_{ij}$.  Some of the eigenvectors correspond
to ``critical'' observables like $M$ or $E$, and the rest 
are ``trivial.''  They can be classified either by inspecting
the probability distributions $p(V_\alpha)$ and $p(V_\alpha,V_\beta)$,
or by looking at the finite volume behaviour of the eigenvalues.  For
example, the $M$- and $E$-like eigenvalues diverge with the critical
exponents as ($L$ is the length of the lattice)
\be
  \lambda_M  \propto L^{3+\gamma/\nu}\,, \h
  \lambda_E  \propto L^{3+\alpha/\nu}\, .
\ee
The ``trivial'' eigenvalues diverge as $L^3$.  A somewhat related
method has been used to study the critical behaviour of the 4d U(1)+Higgs
model \cite{alonso}.

The $M$-like and $E$-like eigenvalues are shown in \fig\ref{fig:chis}
($\chi_a = \lambda_a/L^3$);  the other eigenvalues do not show any critical
behaviour. The results are consistent with the Ising model ones.  
% Also, the joint probability
% distribution $p(V_M,V_E)$ is much closer to the Ising one
% than the distributions shown in \fig\ref{fig:density}.
The positive value of $\alpha$ clearly excludes O($N$) models with
$N\ge 2$ ($\alpha < 0$) and the mean field behaviour ($\alpha = 0$).
%We have also measured the skewness of the $E$-observable $\langle
%E^3\rangle/\langle E^2\rangle^{3/2}$ and the critical exponent $\nu$.
%These are again compatible with the 3D Ising model values.

%%%%%%%%%%%%%%%%%%%%%%%%%%%%%%%%% FIGURE
\begin{figure}[t]

\vspace{2mm}
\hspace*{5mm}\epsfxsize=6cm\epsfbox{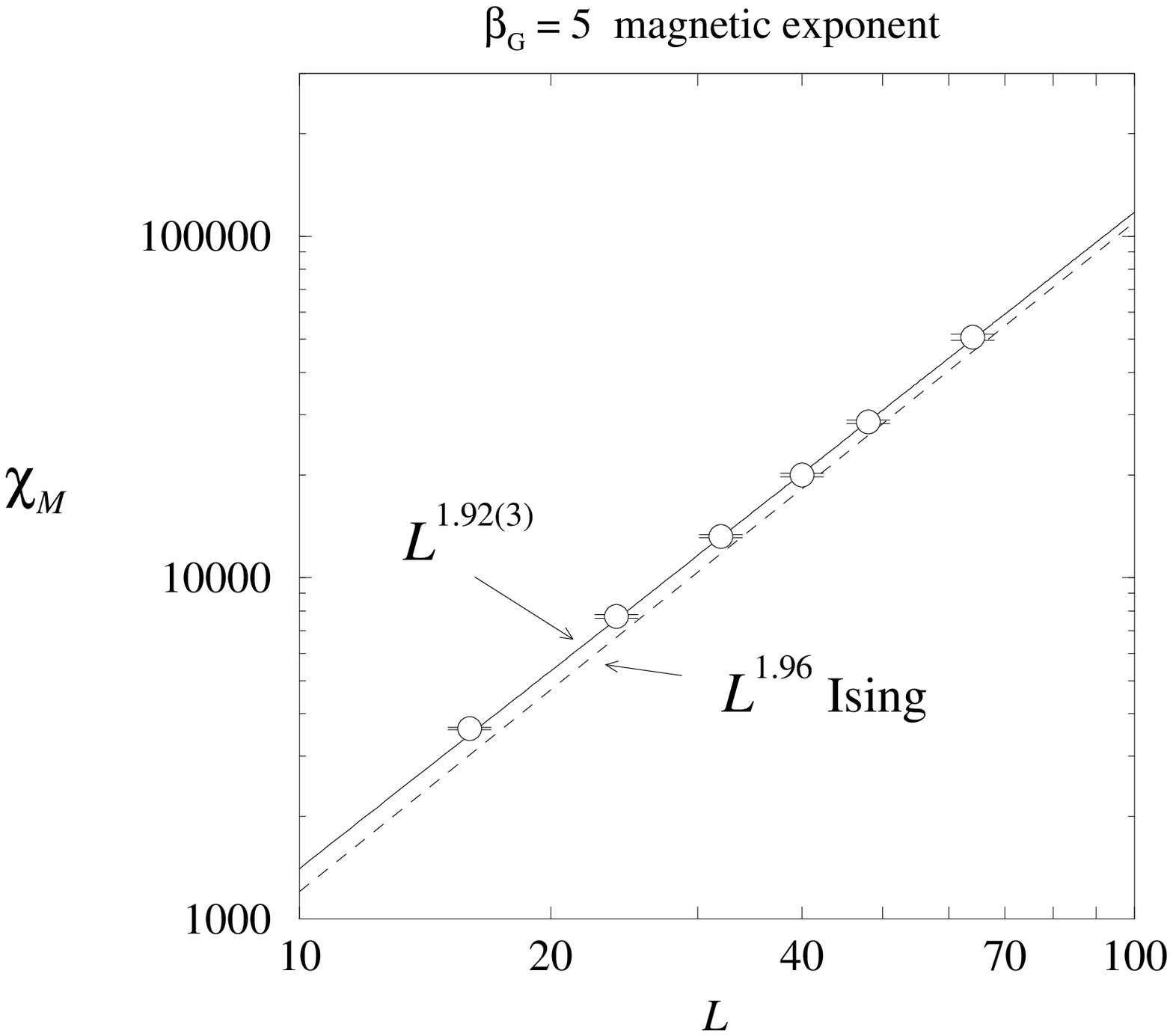}

\hspace*{5mm}\hspace{5mm}\epsfxsize=5.5cm\epsfbox{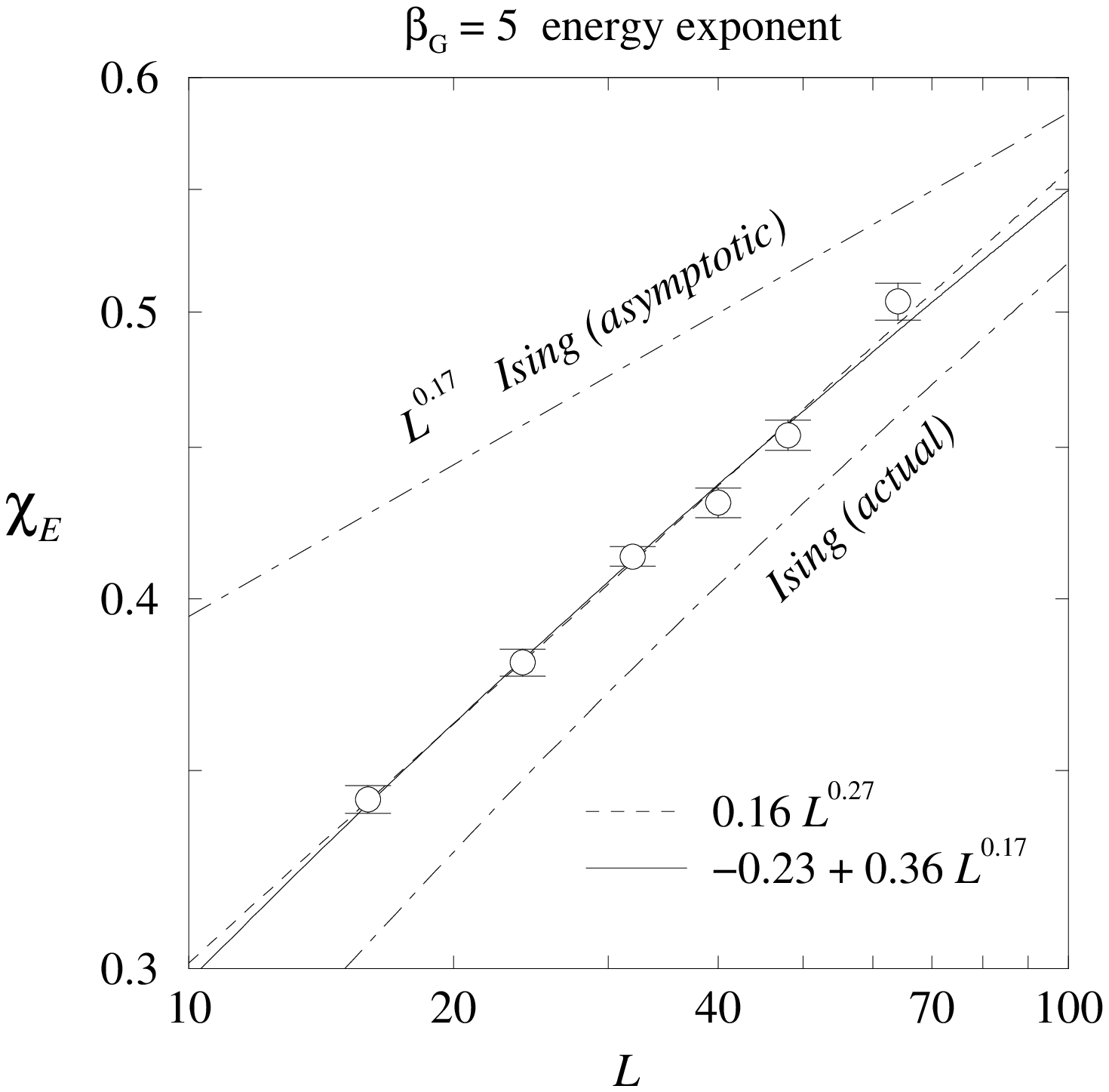}

\vspace{-7mm}
\caption[a]{
The divergence of $\chi_M$ ({\em top})
and $\chi_E$ ({\em bottom}) as a function of the lattice size.
The slope of $\chi_E$ does not agree with the
asymptotic Ising value, but it is 
consistent with the {\em measured\,} \cite{3dising}
Ising model behaviour at these lattice sizes.}
\la{fig:chis}
\vspace{-4mm}
\end{figure}
%%%%%%%%%%%%%%%%%%%%%%%%%%%%%%%%%%%%

%%%%%%%%%%%%%%%%%%%%%%%%%%%%%%%%%% FIGURE
%\begin{figure}[t]

%\epsfxsize=5.5cm\epsfbox{xc_bg.eps}

%\vspace{-6mm}
%\caption[a]{ The infinite volume extrapolations of $x_c$ as a function
%of $\beta_G$.  G\"urtler et al refers to \cite{gurtler}.  The
%$O(a)$ -correction has been calculated in ref.~\cite{moore}.}
%\la{xcrit_betag}
%\end{figure}
%%%%%%%%%%%%%%%%%%%%%%%%%%%%%%%%%%%%%

We have performed the analysis also with 4 instead of 6 operators.
The results remain stable, although in some cases small
deviations from the Ising behaviour begin to appear.  This shows both
the robustness of the method and the importance of including a large
enough number of operators in the analysis.

The Ising-type universality seen in the SM is quite compatible with the
lack of a true order parameter.  The effective $\pm M$-symmetry is not a
symmetry of the action, but it is dynamically generated at the
critical point.  In this respect the system is completely analogous to
the critical point in liquid-vapour transitions.

In the simulations above the lattice spacing was fixed through
$\beta_G = 4/g_3^2a = 5$.  We have also located the critical point at
$\beta_G = 8$, and G\"urtler \etal~\cite{gurtler} have published
results at $\beta_G = 12$ and 16.  This allows us to calculate the
continuum limit extrapolation of the critical point, with the result
$x_c = 0.0983(15)$.  In the Standard Model this 
corresponds to $m_H = 72(2)$\,GeV\@.  Since the experimental lower
limit is $\sim 88$\,GeV, this excludes the existence of the SM
phase transition.  Nevertheless, a first order phase transition is
still allowed in several extensions of the SM\@; most notably, it can
occur in the Minimal Supersymmetric SM.

\end{document}